\documentclass[english,onecolumn]{revtex4}
\usepackage[T1]{fontenc}
\usepackage[latin9]{inputenc}
\usepackage{textcomp}
\usepackage{graphicx}
\usepackage{babel}
\usepackage[T1]{fontenc}
\usepackage[latin9]{inputenc}
\usepackage{graphicx}
\usepackage{babel}

\usepackage{subfig}
\usepackage[subfigure]{tocloft} % no number for Vita in ToC

\usepackage{fancyvrb}
 \usepackage{mathrsfs}
\usepackage{amsmath}
\usepackage{amssymb}
\usepackage{wasysym}
\usepackage{graphicx}
\usepackage{esint}
\usepackage{dcolumn}

\begin{document}
\title{Possible Bubbles of Spacetime Curvature in the South Pacific}
\author{Benjamin K. Tippett}

\email{barn@titaniumphysics.com}

\affiliation{Department of Mathematics and Statistics  \\ University of New Brunswick\\ Fredericton, NB, E3B 5A3 \\ Canada}

\begin{abstract}
In 1928, the late Francis Wayland Thurston published a scandalous manuscript
in purport of warning the world of a global conspiracy of occultists.
Among the documents he gathered to support his thesis was the personal
account of a sailor by the name of Gustaf Johansen, describing an
encounter with an extraordinary island. Johansen's descriptions
of his adventures upon the island are fantastic, and are often considered
the most enigmatic (and therefore the highlight) of Thurston's collection of
documents.

We contend that all of the credible phenomena which Johansen described
may be explained as being the observable consequences of a localized
bubble of spacetime curvature. Many of his most incomprehensible statements
(involving the geometry of the architecture, and variability of the
location of the horizon) can therefore be said to have a unified underlying
cause. 

We propose a simplified example of such a geometry, and show using
numerical computation that Johansen's descriptions were, for the most
part, not simply the ravings of a lunatic. Rather, they are the nontechnical
observations of an intelligent man who did not understand how to describe
what he was seeing. Conversely, it seems to us improbable that Johansen
should have unwittingly given such a precise description
of  the consequences of spacetime curvature, if the details of
this story were merely the dregs of some half remembered fever dream. 

We calculate the type of matter which would be required to generate such
exotic spacetime curvature. Unfortunately, we determine that the required
matter is quite unphysical, and possess a nature which is entirely alien to all of the experiences of human science. 
 Indeed, any civilization
with mastery over such matter would be able to construct
warp drives, cloaking devices, and other exotic geometries required
to conveniently travel through the cosmos. \end{abstract}
\maketitle

\section{Introduction}

\subsection{The Personal Accounts of Thurston, Dyer, and Johansen}

The most wonderful thing in the world, in our opinion, is the ability
of the human mind to correlate many seemingly unrelated pieces of
information into a jubilant whole. We are born ignorant, imprisoned
by the islands of our personal experience; but intelligence, logic,
and diligent study are like glorious seaworthy vessels which allow
us to travel boundless and brilliant oceans. The great ambition of
science is the piecing together of dissociated knowledge to create
hard tempered theories, and then the bravely facing of their philosophical
implications in order to begin the process anew. In this way we have climbed
towards the brilliant truth, and have lifted our human state into
the glory of an age of enlightenment. 

Let us begin with a discussion of the source material our research is based upon,
lest we find our own work tainted by the stigma associated with it. We would like to make it clear that we in no way endorse or condone their occultist perspective. 

In 1928, a manuscript written by the late Francis Wayland Thurston
was published \cite{Call}, relating his conclusions regarding
an investigation begun by his late uncle, Dr. George Angell. These
findings, taken in conjunction with the reported findings of Dr. William Dyer \cite{mnts}
from his 1930 expedition to the Antarctic continent, paint an incredible
picture. As they describe it, somewhere in the oceans of the south
pacific dwell a dormant race of ancient cyclopean monsters. 

The respective works of Angell and Dyer were received with an appropriately generous
quantity of incredulity. The published accounts of Dyer's adventures
in Antarctica have been dismissed as being due hallucinations incurred
by a high-altitude cerebral edema. Thurston's manuscript, on the other
hand, has often been interpreted as being the creative work of a paranoid
mind in the grip of some tragic delirium. The year preceding his death
(and the publication of his manuscript) was spent in the clutches
of a manic obsession which exhausted him mentally and physically. 

It is not a remarkable thing that these two men told similar stories.
As is apparent from their respective accounts, both Dyer and Thurston
were familiar with (what has been referred to as) the \emph{Cthulhu
cult. }If we are to trust that Dyer was indeed hallucinating, knowledge
of the mythology of this cult surely served as fodder for his imaginings.
Alternatively, since this cult was the focus of Thurston's mania, it
should not be surprising that most of the conclusions he drew
were consistent with its mythology. Furthermore, the two stories taken together
corroborate each other no more than separate sightings of the Virgin
Mary could be used to confirm the doctrines of Catholicism. A skeptic
would properly argue that since both delusions were grown from the
same source material, it is natural that they would be consistent
with one another. Additionally, Dyer was a professor at Miskatonic
University, a haven (at the time) for occultist academics
among whom Thurston's manuscript was received with much acclaim. It
is, therefore, not unreasonable to guess that Dyer may have been directly familiar with Thurston's writings before he set out on his fateful
expedition. 

Notable among the objects of interest which
Thurston had collected to support his thesis is the  personal record of Gustaf Johansen, a Norwegian Sailor. Johansen's
record, accurately summarized in Thurston's manuscript, described
the fate of \emph{the Emma}, a Schooner from New Zealand. Johansen was
the second mate of the Emma, and has been described in the
Sydney Bulletin as being sober and a man of some intelligence \footnote{Note:
Johansen's personal testament 
is currently archived in the rare books section at the main library
at Miskatonic University. While we have not yet been able to inspect
them ourselves, academic consensus has it that Thurston's summary
is consistent with it in most details \cite{Joshi}. Thus, the specific descriptions we will be referring to are Thurston's, and not Johansen's directly.}.

Johansen describes an adventure --occurring between March 22 and April
12, 1925-- wherein he and his crew-mates first fought a party of pirates
and then discovered an uncharted island where all of the crew but he
met their fate. The loss of the Emma and the demise of her crew are
well documented; and scholars investigating the Johansen document
have confirmed that it was written in his own hand. Thus, we are convinced of the pedigree of Thurston's documents.

Even if we can trust the paper, to what degree can we trust the words written upon it? On the one hand,  the details of his experience (which he alone survived) are truly
extraordinary and unbelievable. Furthermore, at the time of his rescue Johansen was delirious and out of his wits. On the other hand, the physical
evidence found on Johansen's person and the circumstantial evidence surrounding the event lend the story a degree of credulity.

We feel that our paper will serve to add more fuel to the fire of this debate, since its purpose is to provide a unified explanation for many of Johansen's seemingly \emph{nonsensical} descriptions.
It is our contention that  most of these details are consistent with the hypothesis that Johansen encountered a region of anomalously curved spacetime. To facilitate
our argument, we will propose a simple spacetime geometry which possesses
all of the relevant qualities. We will then, point by point, use this
geometry to explain Johansen's  enigmatic experience and justify his words.

\subsection{Non-Euclidean Geometry and Gravitational Lensing}

As Johansen recounts his adventures upon the
uncharted island, he remarks several times about the frightening geometric qualities
of the place. Thurston describes them using the phrase \emph{Non-Euclidean}, although
its intended meaning is not apparent. Generally,
his use of the phrase is interpreted as referring to the architecture
of the buildings upon the island.  We contend that this orthodox interpretation is incorrect. Indeed, most
sophisticated architecture which integrates any curvature of any kind could correctly be
described as being non-Euclidean. Yet, no one who has explored the notable
edifices of Frank Gehry's design or the roman coliseum would describe the experience as nightmarish
or cyclopean.

 We intuit that it was not the walls of the buildings,
but rather the interstitial space itself which
possesses a curvature. The distinction is simple to describe. Orthodox
interpretations have it that Johansen is describing an architecture
which could be drawn as curved lines on a flat page. We contend
that Johansen is describing an architecture comprised of (what
amounts to) straight lines inscribed upon a curved surface (such as as
saddle or a sphere). 

Einstein's theory of general relativity posits that space alone cannot
be curved, since (to put it colloquially) space and time are different directions along the
same surface \cite{wald1984}. Thus, it is natural to begin our investigation with
the hypothesis that it is spacetime itself which carries the curvature.

The curvature of spacetime is associated with many well known effects. The most common of these is poetically referred to as \emph{gravitational
lensing}, wherein the \emph{image} of objects laying beyond a curved region will
become warped and skewed as gravity bends the trajectories of rays
of light \cite{POISSON}. 

One example of Johansen experiencing this effect  is contained in
the following quote (from Thurston's summary):
\begin{quotation}
A great barn-door ... they could not decide whether it lay flat like
a trap-door or slantwise like an outside cellar-door. ... the geometry
of the place was all wrong. One could not be sure that the sea and
the ground were horizontal, hence the relative position of everything
else seemed phantasmally variable.
\end{quotation}
\begin{quotation}
... all the rules of matter and perspective seemed upset.
\end{quotation}
From this description, It is clear that rays of light, which aught
to be straight, are apparently curving in unreliable ways. As they
walk about the island, they see the outside world (and other distant
objects upon the island) as if through a large
fishbowl. Thus, the horizon would no longer be reliably straight, and the sun and
moon would swing wildly through the sky depending on one's position.
The experience would be enough to drive an expert in nautical navigation quite mad. 

The hypothesis that it is spacetime itself which is curved (and not just
the shape of the buildings) allows us to decipher the enigma of the
large door. 
\begin{quotation}
Then Donovan felt over it delicately around the edge, pressing each
point separately as he went. He climbed interminably along the grotesque
stone moulding - that is, one would call it climbing if the thing
was not after all horizontal - and the men wondered how any door in
the universe could be so vast. 
\end{quotation}
It is clear from this passage that the surface of the door is geodesically flat,
but that it is so large in size with respect to the curvature of space
that the image of the far edges are skewed by gravitational lensing.
Thus, the image of the horizontal door will seem curved and inconsistently oriented
with respect to the horizon as you wander about it.

If Johansen's amazing experiences are indeed due to the curvature
of spacetime: in what specific manner, then,  is it curved? While Johansen's story lacked useful quantitative information, one detail provides
us with a good clue. From Thurston's summary:
\begin{quotation}
... Johansen swears [Parker] was swallowed up by an angle of masonry which
shouldn't have been there; an angle which was acute, but behaved as
if it were obtuse.
\end{quotation}
By this, we deduce that at any time the three dimensional space in
and around the island will be curved hyperbolically (we provide a more thorough analysis in  Sec.\ref{sub:...-an-angle}).

 In order
to further explore the possibility that Johansen has experienced a bubble of curved spacetime, let us introduce a simple spacetime
geometry, and discuss how its specific features can account for what
Johansen has seen.

\section{A Bubble of Localized Spacetime Curvature}

\subsection{Model of Spacetime Curvature \label{sub:Model-of-Spacetime}}

Tests at cosmological scales seem to indicate that our universe bears no
large scale 3-dimensional spatial curvature: it seems to be flat \cite{WMAP}. Let us therefore
conjecture a spacetime consisting of two regions: a spherical internal region with a hyperbolic spatial curvature, and a flat external region. We will refer to this type of geometry as a \emph{bubble}.

Consider the family of spacetime metrics which have the following
line element in geometrized coordinates:
\begin{equation} \begin{split}
ds^{2}=-\left[1-H(r,t)\left(1-\frac{1}{(t+1+\frac{W}{b})^{2}}\right)\right]dt^{2} &+2\sinh(r)(1-H(r,t))dtdr 
\\+ & dr^{2}+\sinh^{2}(r)d\Omega^2
\end{split}
\end{equation}
where $H(r,t)$ serves to transition between the two regions by satisfying $H\approx1$ inside of the spacetime bubble,
and $H\approx0$ outside of the bubble. In order to generate a smooth (but efficient)
transition between the interior and the exterior we will be setting:
\[
H(r,t)\equiv\frac{1-\tanh(A\left(\sqrt{b^{2}t^{2}+r^{2}}-W\right))}{2}\quad,
\]
where the parameter $b$ determines how elongated the bubble is in time, $W$
determines the maximum width of the bubble in space, and the size of $A$ determines the severity
of the transition between the inside and the outside of the bubble.
The coordinate radius which can be described as the edge of the bubble
is then
\begin{equation}
r=\sqrt{W^{2}-b^{2}t^{2}}\;.\label{eq:radius_of_bubble}
\end{equation}

The evolution of the spacetime bubble can be described as follows. At 
time  $t=-\frac{W}{b}$, the bubble of curvature expands
from the point $r=0$, and continues expanding until it reaches its
maximum width $r=W$ at time $t=0$. The radius bubble then begins
contracting, and shrinks down to a point $r=0$ at time $t=\frac{W}{b}$. 

We imagine that perhaps a bubble of curvature of a nature similar
to this one has enveloped  Johansen's uncharted island. It is not clear
from his description \emph{at what stage} along the bubbles' evolution Johansen encountered it, whether and at what rate the bubble was expanding or contracting. Nor is it clear how wide it is, or the degree to which the interior is spatially curved.

\subsection{General qualities of the geometry within the bubble \label{sub:General-Properties-inside}}

Since $H\approx1$ inside of the bubble, the interior geometry can
be approximately described using the metric:
\begin{equation}
ds^{2}=-d\tau^{2}+dr^{2}+\sinh^{2}(r)d\Omega\label{eq:natural_coordinates}
\end{equation}
whose natural (affine) timelike coordinate is related to the exterior
time coordinate according to  transformation $\tau=\ln(t+1+\frac{W}{b})-\frac{W}{b}$.
An observer inside the bubble will therefore see time outside the
bubble pass at an exponential rate. Additionally, in this coordinate
system, constant time hypersurfaces will manifest a constant negative
(hyperbolic) curvature. 

The stress energy tensor on the inside of the bubble can be approximated:
\[
T_{b}^{a}\approx\frac{1}{8\pi}\left(\begin{array}{cccc}
3 & 0 & 0 & 0\\
0 & 1 & 0 & 0\\
0 & 0 & 1 & 0\\
0 & 0 & 0 & 1
\end{array}\right)\;.
\]

\subsection{General qualities of the geometry outside of the bubble \label{sub:General-Properties-Outside}}

Since $H\approx0$ outside of the bubble, the exterior geometry can
be approximately described using the metric: 
\[
ds^{2}=-dt^{2}+2\sinh(r)dtdr+dr^{2}+\sinh^{2}(r)d\Omega
\]
which is merely the (flat) Minkowski spacetime in hyperbolic coordinates:
\[
ds^{2}=-d\sigma^{2}+dR^{2}+R^{2}d\Omega
\]
\[
R=\sinh(r)\;,\;\sigma=\cosh(r)+t\;.
\]
Therefore, an observer sitting a sufficient distance outside the bubble
will feel no curvature.

\subsection{Our choice of parameters}

In the subsequent sections, we will be investigating the various features
of this bubble of curvature using numerical computation.
In order to do so we must assign some values to the free parameters: $A, \; b ,\; W$.  Let us specify: $b=\frac{1}{40}$, which gives the
bubble the shape of an elongated prolate spheroid in spacetime; $A=10$,
which ensures a relatively sudden transition between the interior
(hyperbolic) and the exterior (flat); and $W=5$, which sets
the maximum radius of the bubble to be at the coordinate $r=5$. While
it is possible to choose a value for the $b$ parameter which
would allow the bubble to persist for aeons, it is more convenient for
our current purposes (numerical computation) to deal with a bubble
whose temporal and spatial dimensions are comparable.

\section{Using the spacetime geometry to interpret the details of Johansen's
story}

There are a variety of enigmatic and elusive descriptions in Thurston's
summary of Johansen's testimony. Historical interpretations have made
little sense of them, and consensus has it that they mostly serve
to give the tone of Johansen's anecdote a mysterious and alien quality. 

In context of a spacetime bubble, however, these descriptions seem both technically detailed and reasonable. 

\subsection{\emph{``An angle which was acute, but behaved as if it were obtuse."}
\label{sub:...-an-angle}}

From Thurston's Summary of Johansen's notes:
\begin{quotation}
... Johansen swears [Parker] was swallowed up by an angle of masonry which
shouldn't have been there; an angle which was acute, but behaved as
if it were obtuse.
\end{quotation}
This description is confusing because while it is sparse on technical detail, 
it is also dense with information. Thurston and Johansen are clearly not
well versed in the vocabulary of non-Euclidean geometry --we would not expect them to be-- and they leave us with the task of deducing the
underlying truth. 

Evidently, Johansen is describing an event wherein his crewmate passed between
a straight beam (or column, buttress, or wall) and the ground (or some other reliably flat
surface). The beam must be straight because there would be no meaning
in mentioning the angle of the masonry if it were curved. Thus,
we interpret his saying that the ``angle ... was acute but behaves as if it
were obtuse\emph{''} as meaning that he was surprised at the ample
quantity of interstitial volume between beam and the ground. We can imagine him
glancing at the angle at which the beam met the ground, recognizing
its shallownewss, and then being surprised that his crewmate was able to fit
his body between the two surfaces. Since it is reasonable
to suppose that the respective surfaces have little inherent curvature
themselves, the anomalous feature must lie not in the
masonry itself but rather in the nature of the interstitial space (i.e. the space itself must have curvature).

We are used to \emph{flat} space or, rather, \emph{space bearing no intrinsic
curvature}. The relationships and laws detailing the lengths of lines,
areas, and volumes are those which were compiled in the ageless writings
of Euclid, and thus, a flat space is commonly described as being \emph{Euclidean}.
 In contrast, all geometric rules relating to lines, areas,
and volumes on curved surfaces are described as being \emph{Non-Euclidean.}

In Euclidean geometry, the triangular area subtending two straight lines which
meet at an acute angle $\theta$ will be $A_{flat}=\frac{1}{2}\ell^{2}\tan(\theta)$
where $\ell$ is the length of the line lying at the base of the triangle
 (as Euclid described, two such lines will meet at one and only
one point).

Contrast this to \emph{elliptic} spatial geometry, or \emph{space bearing
a constant positive curvature}. The simplest way to understand elliptic geometry is to imagine drawing lines and painting in areas upon the surface
of a large sphere. In such a situation locally \emph{straight lines} lie as 
\emph{great circles}  upon the sphere (like the equator, or lines of longitude). Thus,
any two straight lines are destined to intersect in two places, and
the area subtended between two such straight lines must satisfy $A_{elliptic}<\frac{1}{2}\ell^{2}\tan(\theta)$.
If the three dimensional space of Johansen's island were to have an
elliptic spatial geometry, he would have been surprised to discover
how little area would be available for Parker to fit between the beam
and the ground, given their angle of intersection. 

The third possible spatial geometry is called  \emph{hyperbolic }spatial
geometry, or \emph{space with a constant negative curvature}. The simplest
way to understand such a system is to imagine drawing lines and painting in
areas upon the surface of a large saddle. In such a situation, locally
\emph{straight lines }are set to lie as hyperbola upon the saddle. In hyperbolic geometry, two lines which cross at a point
will spread apart at a much faster rate than they would in Euclidean
geometry. The area subtended between two such straight lines must
satisfy $A_{hyperbolic}>\frac{1}{2}\ell^{2}\tan(\theta)$. Thus, we contend that
the three dimensional space of Johansen's island must have had a hyperbolic
spatial geometry, justifying his surprise at the
area  between the beam and the ground.

\subsection{\emph{``The relative position of everything else seemed phantasmally
variable."}}

\begin{figure} \centering
             \subfloat[Tracing the paths of individual light rays, we note that their trajectories
will be curved while inside the spacetime
bubble, they will diffract a small amount as they cross the boundary
between the regions, and  they will move in \emph{straight }lines
once beyond the bubble. ]{\label{fig:raytrace} \includegraphics[trim=6mm 5mm 2mm 4mm,clip,scale=0.4]{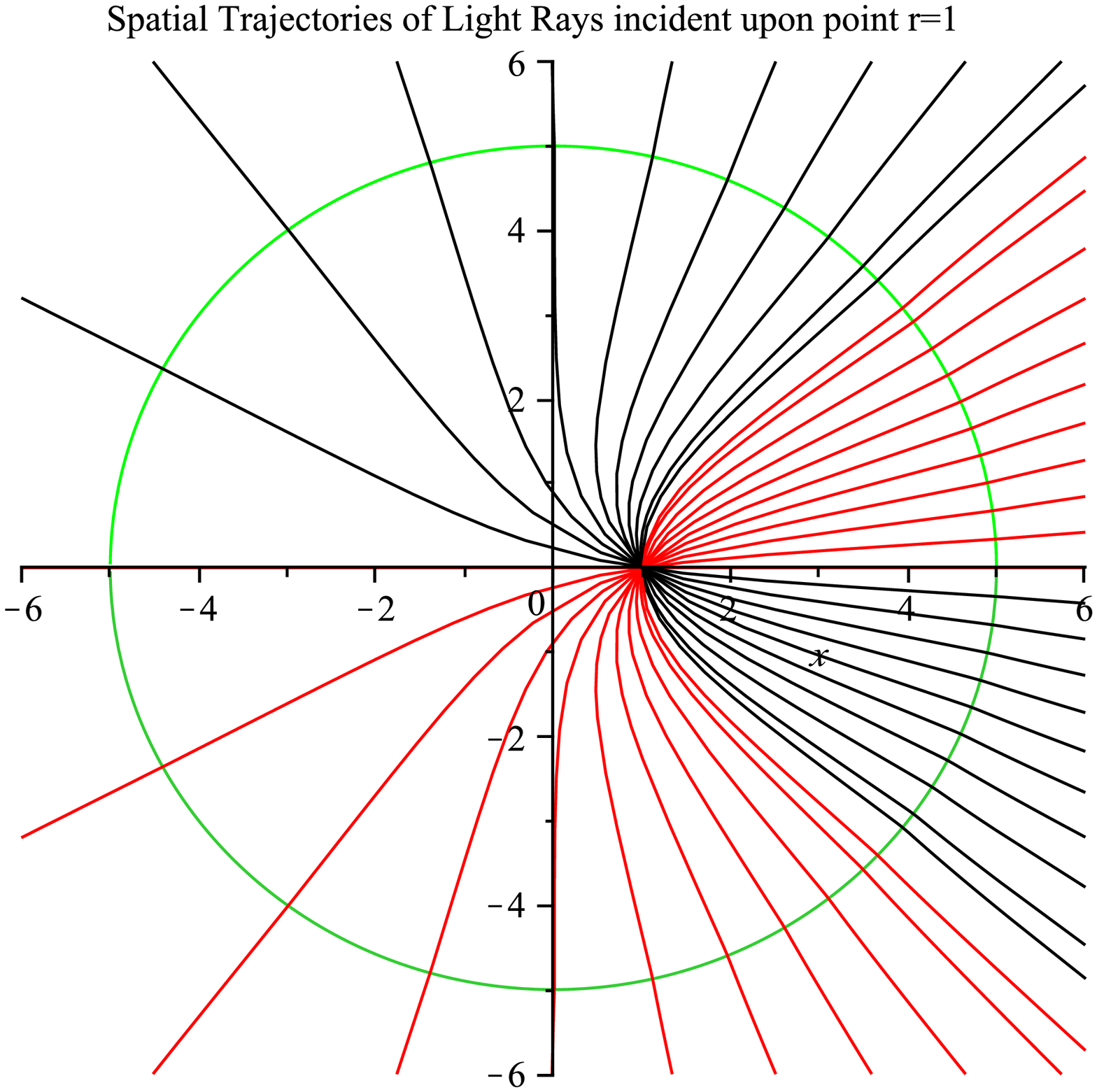} }  $\;$
         \subfloat[The relationship between the actual angular coordinate of an object, and its apparent angle. This can be used to visualize how gravitational lensing will affect the visual field for an observer on the island. ]{\label{fig:anglein} \includegraphics[trim=2mm 1mm 0mm 0mm,clip,scale=0.4]{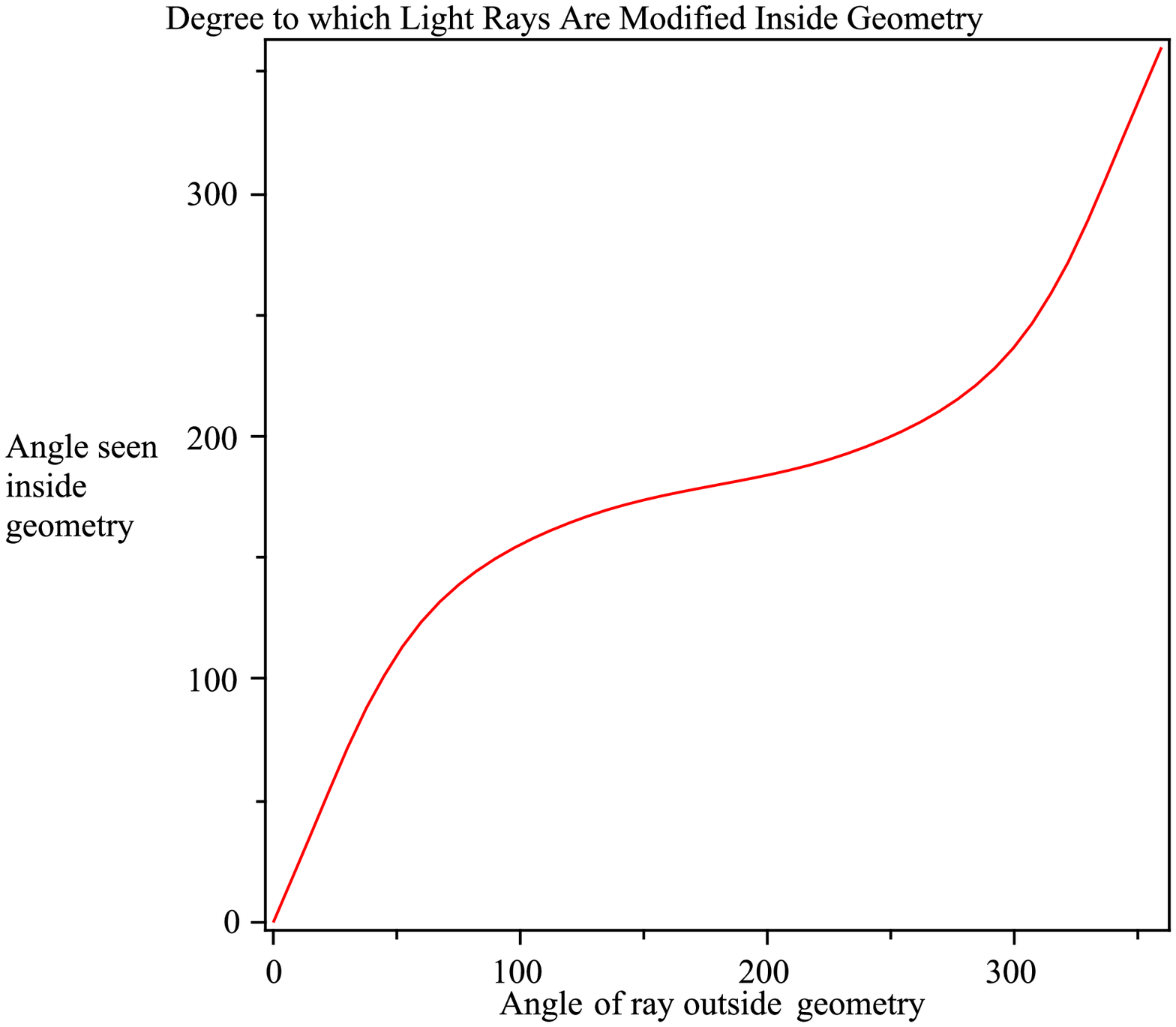}}
     
          \caption{Through ray tracing, the appropriate modifications to the angular dimension and distribution of objects in the visual field can be computed.  }
\end{figure}

\begin{figure} \centering
{ \protect\includegraphics[trim=0.5mm 1mm 9mm 0mm,clip,width=0.3\textwidth]{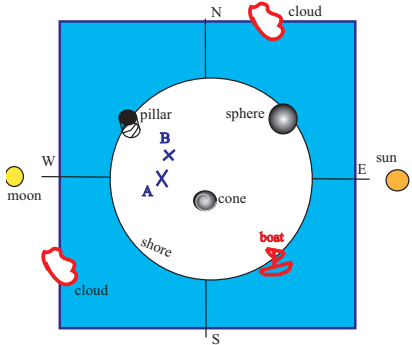}}
          \caption{Visual panoramas of this island  from positions A  and B  will be generated in Fig.\ref{fig:Aplace} and Fig.\ref{fig:Bplace}, respectively. \label{fig:crazytownx}}
\end{figure}

\begin{figure} \centering

             \subfloat[Unlensed  A panorama]{\label{fig:Aun} \includegraphics[trim=0mm 1mm 0mm 0mm,clip,width=1.25 \textwidth,angle=90]{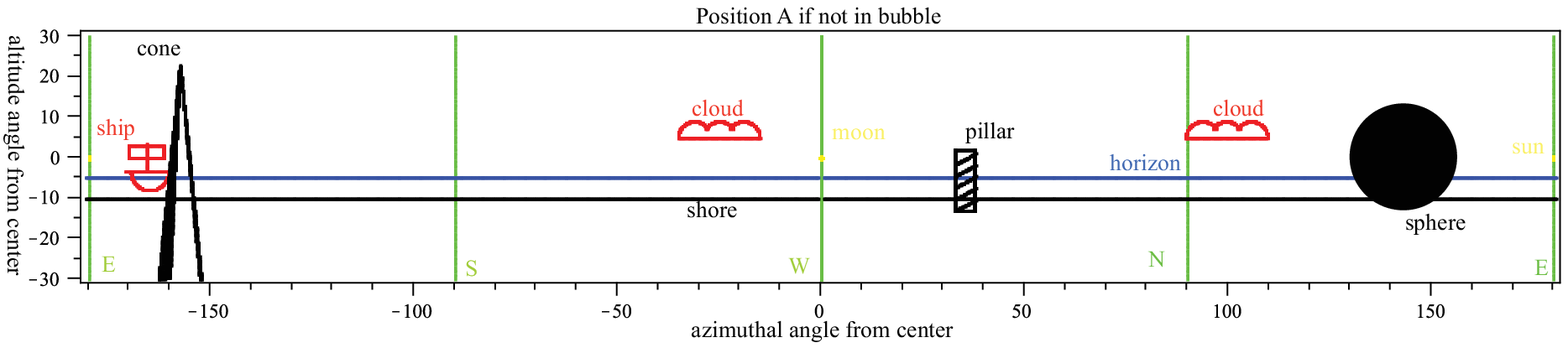} } $\; \; \;$
         \subfloat[Lensed panorama from A ]{\label{fig:Bun} \includegraphics[trim=0mm 1mm 0mm 1mm,clip,width=1.27 \textwidth,angle=90]{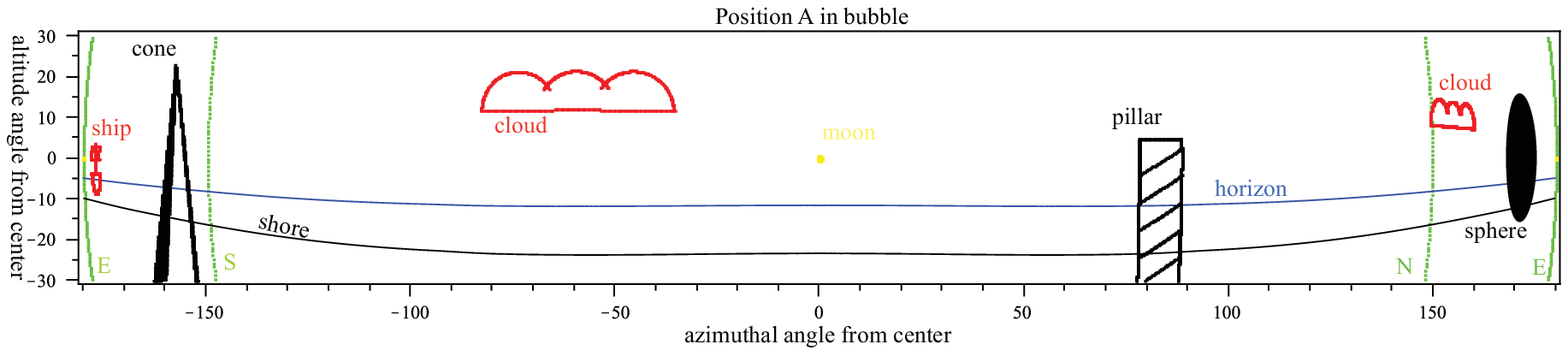} }
         \subfloat[Lensed panorama farther out from A]{\label{fig:Alen} \includegraphics[trim=0mm 10mm 0mm 18mm,clip,width=1.215 \textwidth,angle=90]{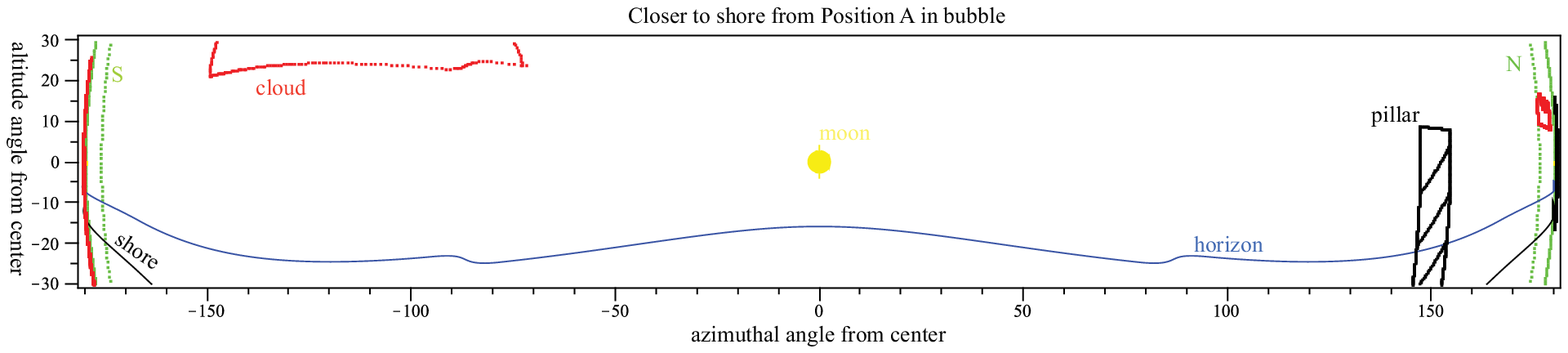} }
     
          \caption{Visual panorama with and without gravitational lensing from A and from a wider radius than A on the island. Note how dramatic the effect of lensing is, and how more pronounced the lensing becomes as we approach the shore.  \label{fig:Aplace}}
\end{figure}

\begin{figure} \centering

             \subfloat[Unlensed  B panorama]{\label{fig:Aun2} \includegraphics[trim=0mm 1mm 0mm 0mm,clip,width=1.25 \textwidth,angle=90]{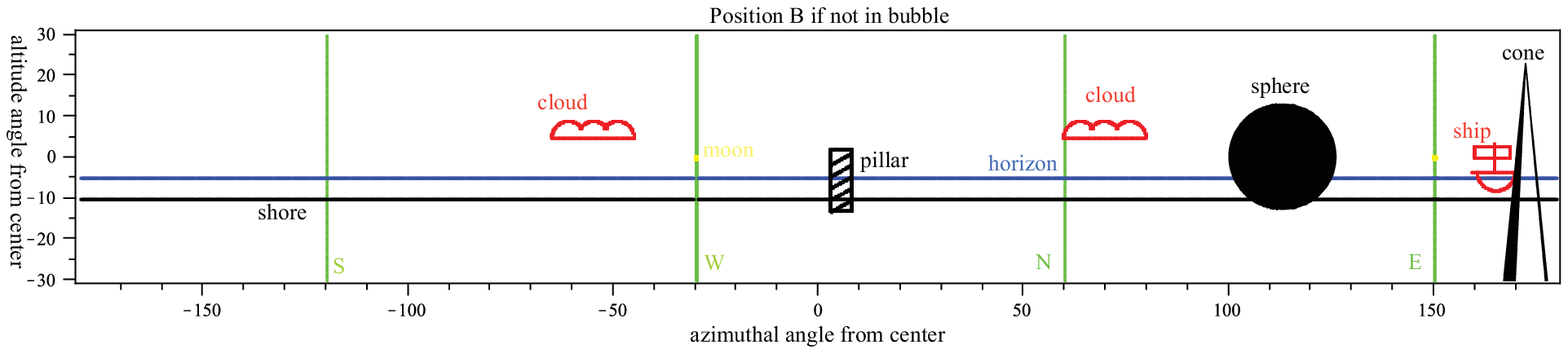} } $\; \; \;$
         \subfloat[Lensed panorama from B ]{\label{fig:Bun2} \includegraphics[trim=0mm 1mm 0mm 1mm,clip,width=1.265 \textwidth,angle=90]{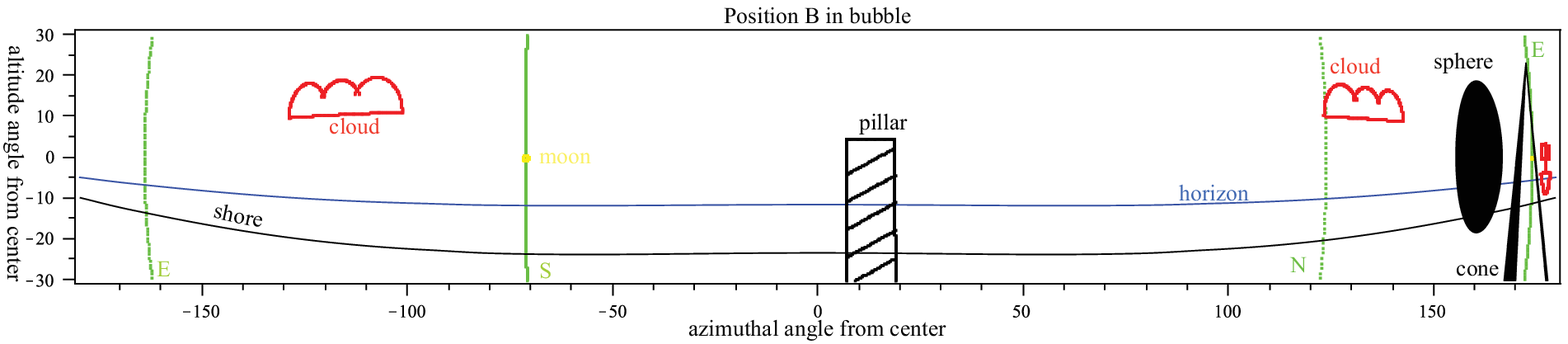} }
         \subfloat[Lensed panorama farther out from B]{\label{fig:Alen2} \includegraphics[trim=0mm 0mm 0mm 0mm,clip,width=1.25 \textwidth,angle=90]{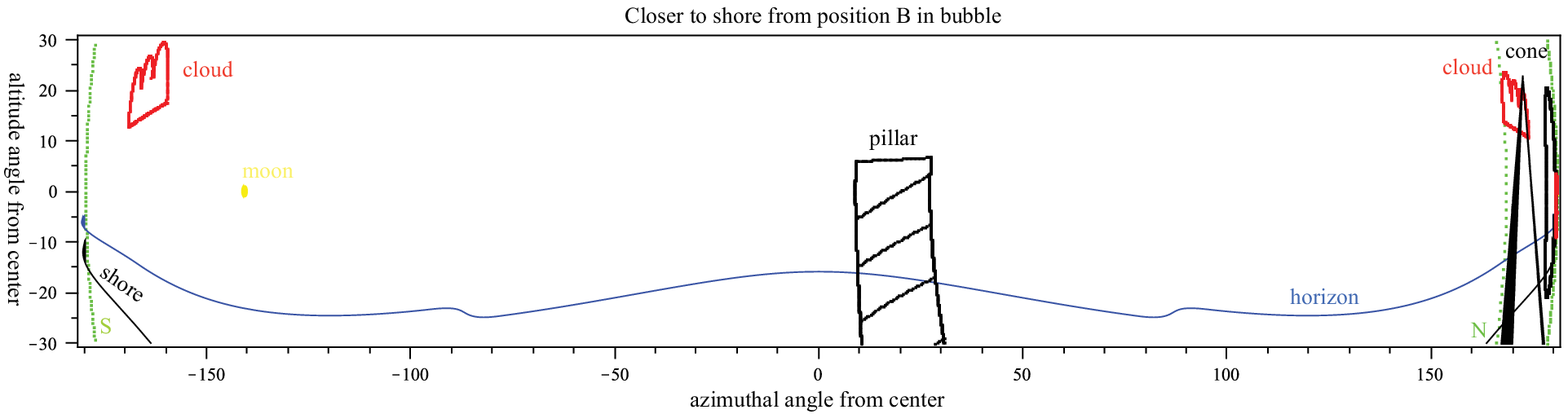} }
     
          \caption{Visual panorama with and without gravitational lensing from position B and from a wider radius than B on the island. Note how dramatic the effect of lensing is, and how more pronounced the lensing becomes as we approach the shore.   \label{fig:Bplace}}
\end{figure}

Thurston's summary of Johansen's testimony contain tantalizingly enigmatic
descriptions of objects on the island. 
\begin{quotation}
A great barn-door ... they could not decide whether it lay flat like
a trap-door or slantwise like an outside cellar-door. As Wilcox would
have said, the geometry of the place was all wrong. One could not
be sure that the sea and the ground were horizontal, hence the relative
position of everything else seemed phantasmally variable.
\end{quotation}
We interpret these words as meaning that Johansen examined the door
and upon close inspection found that it was flat at all points upon
its surface. Inspecting it from farther away, or comparing it to objects
in its periphery, however, seemed to present an inconsistent picture
regarding its curvature and dimension. While this effect may
seem frighteningly alien to a sailor; it is  common enough
in the field of spacetime geometry. Extracted from their adventurous context,
his words might very well be describing the equivalence principle and gravitational
lensing!

In a curved spacetime, rays of light no longer travel upon reliably
\emph{straight} trajectories, and so the visual field is no longer
a reliable tool to describe the relative positions of two objects
or the flatness of an extended object. Let us demonstrate how an observer
immersed in our spacetime bubble would make observations which are
consistent with Johansen's.

We suppose (for computational purposes) that an observer is placed
at $t=0$, $r=1$, $\theta=0$ within our spacetime bubble. How would the outside
world appear to him? While our regular experiences insist that light
rays should travel in straight lines (or alternatively, along trajectories
of unvarying angle); light rays in a curved spacetime will travel along
curved paths. The trajectories of light beams which 
intersect his position can be calculated (see Fig.\ref{fig:raytrace}),
and from repeatedly performing this calculation we can determine how the field of vision for an observer
in this position will be modified.

This procedure allows us to determine the relationship between the angle of a ray
as it intersects our observer and the angle the of ray as
it travels outside the bubble (See Fig.\ref{fig:anglein}).
Thus, if we know the angular size subtended by objects at very large
distances outside of the bubble, as well as their location in the
field of vision we can determine how their appearance will be distorted
by the spacetime curvature.

Let us demonstrate the effect of this gravitational lensing by considering a few specific examples (See Fig.\ref{fig:crazytownx} for reference). Suppose that we have two observers, denoted A and B, and suppose that they
are standing upon an island at the center of a spacetime bubble. Observer A is at radius $r=1$ and is located due west of the middle of the bubble. Observer B is at a radius $r=1$ and is located $30^{o}$ north of west from the center. Using numerical computation, we will generate effective panoramas as seen by observers A and B, and for observers at $r=2$ along the same rays as observers A and B. Thus, we will be able to demonstrate how the lensing effect will change as we move about the island.

 For dramatic effect, there are
three geometrical objects placed upon the island along with our observers: a great pillar
and sphere which lie on the periphery of the island (to the northwest and northeast, respectively), and a tall
cone which lies closer to the observers. Beyond the island's shore and out to the horizon lies the ocean, upon and above which floats a ship and some clouds. We will also
suppose that the sun lies in the east $5^{o}$ above the horizon,
and that the gibbous moon lies in the west $5^{o}$ above the horizon
(the width of both the sun and moon subtend $\frac{1}{2}^{o}$). 

The cone lies near the observers and its image will encounter little lensing; whereas all other objects are far from the observers and their images will feel the full effect of the spacetime geometry.

Fig.\ref{fig:Aun} and Fig.\ref{fig:Aun2} show panoramas of the island as seen from positions A and B
if there were no spacetime bubble (and no lensing). The cardinal
directions have also been drawn in.  The range of altitudes drawn in
each panorama is between $+30^{o}$ and $-30^{o}$ to keep the rendering visually accurate.
 Both panoramas
are centered upon the respective directions closest to the edge of the island (and the bubble). Let us refer to this central ray the \emph{center}. Thus, the center of the panorama for observer A lies directly west; and the center for observer B lies $30^{o}$ north of west.

Fig.\ref{fig:Bun} and Fig.\ref{fig:Alen}  show panoramas of the island inside a bubble from position
A and from a position closer to the shore than A (but sharing its panoramic center). The differences between these
images and Fig.\ref{fig:Aun} are attributed to gravitational lensing. Note how much more pronounced the effect becomes as we move away from the center of the island and towards the shore (Fig.\ref{fig:Alen}). We should note that in Fig.\ref{fig:Alen}, the gibbous moon laying along the center of the panorama has an apparent area which is 100X its usual area. 

Fig.\ref{fig:Bun2} and Fig.\ref{fig:Alen2}  show panoramas of the island inside a bubble from position
B and from a position closer to the shore than B (but sharing its panoramic center). The differences between these
images and Fig.\ref{fig:Aun2} are attributed to gravitational lensing. Note how much more pronounced the effect becomes as we move away from the center of the island and towards the shore (Fig.\ref{fig:Alen2}). Note how in Fig.\ref{fig:Alen2}, the shape of the tower becomes elongated and warped. It's no longer apparently vertical, and possesses no right angles.

Surely, these effects  are consistent with Johansen's observation that:
\begin{quotation}
One could not be sure that the sea and the ground were horizontal,
hence the relative position of everything else seemed phantasmally
variable.
\end{quotation}

One should also notice the dramatic difference
between the panoramas seen by observers A and B, who lie a similar distance from the middle of the bubble (Fig. \ref{fig:Bun} and Fig. \ref{fig:Bun2}, respectively). The relative angular distances between objects varies much more wildly than they do in the unlensed panoramas (Fig.\ref{fig:Aun} and Fig.\ref{fig:Bun}). Thus, as an observer
wanders about the island, they would see ordinarily stationary objects in the background moving about wildly, and would argue that the laws of perspective no longer apply.

We could characterize the general effects of the gravitational lensing upon a specific landscape in a few ways.

Firstly, the 
area surrounding the center is greatly expanded and takes up a much
larger portion of the sky. As we compare and contrast the pair  Fig.\ref{fig:Bun} and Fig.\ref{fig:Alen}  and the pair Fig.\ref{fig:Bun2} and Fig.\ref{fig:Alen2}, we see that the effect becomes more pronounced as you move towards the edge of the bubble.

Secondly, the images of all other background objects which are farther away from the center of the panorama will be drawn towards the point antipodal to the center. The effect can be quite dramatic. If we consider panoramas Fig.\ref{fig:Alen}  and Fig.\ref{fig:Alen2}, the whole eastern hemisphere (from north to south) seems to be compressed to within $15^{o}$ of the antipodal points.

\subsection{\emph{``From that time till his rescue on the 12th the man remembers
little..."} \label{sec:td}}

One  enigmatic detail of Johansen's adventure which is not
often discussed is the mystery of how Johansen survived the time
between his encounter with the island and his rescue. Consider, from
Thurston's record:
\begin{quotation}
... The Emma, in ballast, had cleared Auckland on February 20th, ...
the ship was making good progress when held up by the Alert on March
22nd... Then, driven ahead by curiosity in their captured yacht under
Johansen's command, the men sight a great stone pillar sticking out
of the sea, and in S. Latitude 47\textdegree{}9', W. Longitude l23\textdegree{}43'...
\end{quotation}
Thus, Johansen's adventure on the island took place within a few days
of March 22. Thurston's summary of Johansen's story after the island
is as follows:
\begin{quotation}
That was all. After that Johansen only brooded over the idol in the
cabin and attended to a few matters of food for himself and the laughing
maniac by his side. He did not try to navigate after the first bold
flight, for the reaction had taken something out of his soul. Then
came the storm of April 2nd, and a gathering of the clouds about his
consciousness. There is a sense of spectral whirling through liquid
gulfs of infinity, of dizzying rides through reeling universes on
a comets tail, and of hysterical plunges from the pit to the moon
and from the moon back again to the pit, all livened by a cachinnating
chorus of the distorted, hilarious elder gods and the green, bat-winged
mocking imps of Tartarus.

...From that time till his rescue on the 12th the man remembers little...
\end{quotation}
It is clear from the testimony of the people who rescued him that when he was discovered,
Johansen was in a state of extreme agitation. From his own recounting of events, it is clear that he was hallucinating. Impossibly, based on Johansen's
timeline, he remained in this state for a period of time lasting from ten days to over two weeks.
Could an ailing individual, one who is paralyzed with delirium, remember to eat,
drink, sleep and otherwise perform the quotidian maintenance which
our bodies require to survive? We are incredulous at the possibility.

Consistent with the spacetime bubble hypothesis, though, is the possibility
that Johansen experienced time dilation. One effect of the curvature of spacetime in the bubble is that, to put it simply, time passes at a slower rate inside of the bubble. 
Thus, it is reasonable that while two weeks elapsed in the outside
world, only a handful of hours or days were experienced by 
Johansen and his crew over the course of their bizarre adventure.

\subsection{Reinterpreting the central myth of the Cthulhu cult}

We risk entering the domain of the fabulist or occultist
in doing so, but the attributes of the bubble spacetime mentioned in Sec.\ref{sec:td}  unexpectedly offer an explanation for an unrelated detail shared in both Dyer and Thurston's stories.

Explicitly, for an observer sitting at the position $r=0$ (based
on Eq. \ref{eq:radius_of_bubble}) we see that the bubble will last
between $t=\pm\frac{b}{W}$. Thus,  the total time elapsed
in the outside world would be $\Delta t=2\frac{b}{W}$.
On the other hand, the natural time coordinates on the inside of the
bubble $\tau$ (see the metric Eq. \ref{eq:natural_coordinates}) will
only see $\Delta\tau=\ln(1+2\frac{W}{b})$ pass. The fact that\emph{
time passes exponentially faster on the outside of the bubble compared
to the inside of the bubble} means that countless aeons may pass on the outside,
while only moments pass on the inside. 

Both Dyer and Thurston make reference to the \emph{Cthulhu cult}
whose central conceit is that a deity named Cthulhu resides in
a city at the bottom of the sea, waiting for his prophesied return to power. The distinction between
this myth and the common \emph{King under the Mountain} myths are
the morbid details of Cthulhu's current state, most
whimsically  described by the Arabic poet Abd-al-Hazred:
\begin{quotation}
That is not dead which can eternal lie, 

And with strange aeons even death may die.
\end{quotation}
Commonly, the myth describes Cthulhu as being neither dead nor alive. 
From Thurston's manuscript
\begin{quotation}
...They could not live. But although They no longer lived, They would
never really die. They all lay in stone houses in Their great city
of R'lyeh, preserved by the spells of mighty Cthulhu for a glorious
surrection when the stars and the earth might once more be ready for
Them. 
\end{quotation}

Of course, (quantum mechanics  and cats aside) the idea that something can be both alive and dead at the same time is physically nonsensical.
 
However, with our more sophisticated vocabulary, we might describe this Cthulhu
as being in a position where it does not feel the passage of time.
As enigmatic as these words seem, they are entirely consistent with
our spacetime bubble model. Anyone sitting at the center of the bubble would seem (to the outside world) unchanging for aeons to observers in the outside world. 

This is not to say that we agree with Thurston and Dyer.

\subsection{The physical feasibility of the geometry}

At this point, one might ask how physically feasible  such a spacetime
bubble could be. The Einstein equation provides a relationship between
the geometry of the spacetime, and the nature and configuration of
matter whose gravitational influence is required to generate the geometry. Now that we have defined a model of the geometry, we can determine whether it can be generated using realistic matter \cite{wald1984}.

While we have already discussed the nature of the matter outside the
bubble (a vacuum, see Sec.\ref{sub:General-Properties-Outside})
and the matter inside the bubble (see Sec.\ref{sub:General-Properties-inside})
the true nature of the requisite matter is subtler, since more careful distributions of matter will be required to transition between the interior and the exterior regions. 

\begin{figure} \centering
\protect\includegraphics[scale=0.5]{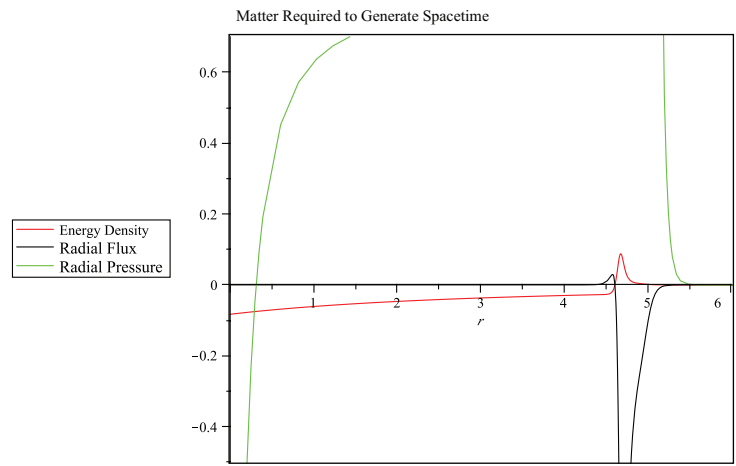}
\caption{\label{fig:energy} The matter required to generate the spacetime geometry. The requirement for negative energy indicates that this geometry cannot be created with any physically recognized matter.}

\end{figure}

To get a sense of the matter at hand, in Fig.\ref{fig:energy} we plot
the radial dependence of various attributes of the matter we require: the
energy density, the radial energy flux, and the radial pressure. Recall that the transition between the interior and the
exterior of the bubble is centered around $r=5.$ 

There are many notable
attributes in Fig. \ref{fig:energy}, but chief among them is the requirement for a
negative energy density within the bubble. Such matter is attributed
with the property of being gravitationally repulsive, and is generally considered
to be unphysical \cite{Hawking}. Thus, the bubble of spacetime geometry which Johansen
explored could not have been made from any type of matter
with which human science is familiar. 

It is worth mentioning, before we conclude this discussion, that many
fascinating spacetime geometries require similar types of exotic matter. The list includes traversable wormholes (and thus time machines \cite{morris}),
 warp drives \cite{alcubierre}, and spacetime cloaking devices \cite{cloak}. Speaking very broadly,  only a 
civilization which has the capacity
to travel through the universe at superluminal speeds would have
the capacity to build Johansen's bubble. 

While we feel that it is distasteful to return to the subject of the occult, due diligence requires that we mention a final coincidental parallel between our findings and a detail common to the works of both Dyer and Thurston. The two authors'
stories both describe the Cthulhu and his ilk
as being a spacefaring race. 

From Thurston's manuscript:
\begin{quotation}
...the Great Old Ones who lived ages before there were any men, and
who came to the young world out of the sky.

...They had, indeed, come themselves from the stars, 

... When the stars were right, They could plunge from world to world
through the sky; 
\end{quotation}
and from Dyer's publication
\begin{quotation}
Another race - a land race of beings shaped like octopi and probably
corresponding to fabulous prehuman spawn of Cthulhu - soon began filtering
down from cosmic infinity...
\end{quotation}

Please do not take this as an endorsement of Thurston and Dyer's theories.

\section{Conclusions}

The most infamous writings of Francis Wayland Thurston and Dr. William
Dyer are enigmatic, and  the bulk of the fantastic details in their manuscripts
have rightly been dismissed.

Dyer was a geologist and a scholar of note, but he himself admits to
keeping the company of occultists during his  tenure at Miskatonic
university. On his disastrous trip to the Antarctic, a cocktail of stress and altitude sickness caused a type of delirium, and the seeds which
the occultists had planted germinated and bloomed into a very strange story
indeed. Consensus among academics is that we should forgive the good Dr.
Dyer his embarrassment, and that posterity aught to focus upon his more serious works.

Francis Wayland Thurston, on the other hand, was clearly in the clutches
of a manic delusion similar to those which possess modern day conspiracy theorists
 and UFO enthusiasts. In the months preceding his
death he collected numerous disparate news clippings, police records
and private testimonials, and then wove them together into a schizophrenic
(and at times very racist) story involving biracial cultists in Louisiana,
Australians, Inuit folk tales, artists with bipolar disorder, senile
academics and an international conspiracy of seamen. The focus of
his manuscript was the founding myth of an occultist organization which  Thurston refers to as the Cthulhu cult.

While Dyer also mentions Cthulhu in his story, no one familiar with the
broader historical context interprets this fact as corroborating Thurston's wild story. Indeed, Thurston's work was published a few years before Dyer's
tragic expedition; and given Dyer's occult-loving companions, it would not come as a surprise if Dyer had
previously heard of Thurston's work.

While consensus has it that Thurston was a fabulist, many of the articles
and events he compiled are reliable and are a matter of public record.
And, as any astronomer who searches for historical supernovae by sifting through records of
chinese astrology  will tell
you, reliable first hand observations can have a value independent of that
of the interpretation within which they are presented. One of Thurston's
compiled articles relays first hand details regarding the demise of a New Zealand schooner (the \emph{Emma}), and the subsequent adventures of its sole
survivor, a Norwegian sailor named Gustaf Johansen. Johansen's personal account
relays his experiences upon an uncharted island in the south pacific. 

Johansen's description of the island, as summarized by Thurston, is enigmatic
and cryptic. Historically, these details have been dismissed as alternatively fabulism
or  artistic license on either Thurston or
Johansen's part. 

The chief conclusion of our research is that  the  bulk of Johansen's
enigmatic observations can be attributed to a region of anomalously curved spacetime, and the the consequential gravitational lensing of images therein. We are able to explain most of Johansen's  observations including the descriptions of weird architectural forms, their inconsistent orientation to the horizon, and the maddeningly unreliable location of objects in the celestial sphere. 

We construct an example of such a bubble of curvature, and demonstrate
that an observer in Johansen's shoes would indeed sound like a lunatic
when asked to describe what they see (Fig.\ref{fig:Aplace} and Fig.\ref{fig:Bplace}). 
As we compare, point-by-point, Johansen's observations with the impressions of the lensed panoramas, we feel that they are consistent.

Our scientific explanation for Johansen's experiences upon the island
allows us to explain other mysteries,  peripheral to his island adventure. For instance,
Johansen was able to survive adrift at sea for nearly two weeks, and he apparently did
so in a state of helpless dementia. If Johansen was, as
we contend, exploring a bubble of curved spacetime, we expect
that he experienced the effect of time dilation. This would shorten
the span of time he would need spend aboard the boat, and in turn, explain his survival.

The only evidence which is currently available to us are those documents which Thurston compiled and published. Thus, any conclusions we draw may only be a statement of likelihood. 

In cases such as these, where one simple explanation serves to decipher and justify so many disparate and enigmatic elements of a story, is not the new-found self consistency compelling? Is not some new degree of credulity granted to the collection of fantastic details which previously seemed so unbelievable?

Conversely, what is the probability that the imagination of a layperson in the 1920's would be able to accidentally describe not just the effects of gravitational lensing but also the consequential anomalous relationship between lines, angles and areas in a curved space? How would he know to provide details to a mystery whose only solution could be time dilation due to curved spacetime? How likely is it that men with no knowledge of modern general relativity would be able to blindly fabricate so many self-consistent details? 

To temper our findings, we concluded our study by asking  whether Johansen's spacetime bubble could be generated with any physically feasible matter.
 As our model demonstrates, an exotic type of matter with which human science is entirely unfamiliar is required for such a geometry to exist. Indeed, this is the very species of energy which is theoretically  
required to build a warp drive or
a cloaking device. Only a people capable of crossing vast
cosmic distances could have constructed Johansen's bubble. Furthermore, whoever constructed such a structure would need command over vast energies, and the capacity to construct edifices on a cyclopean scale. 

As a final point, our model requires that time passes exponentially more quickly on the outside of the bubble than on the inside. Such bubble of non-Euclidean geometry could be used to endure vast aeons of time while the universe outside it grows brittle with age. 

We hesitate to give credence to Dyer and Thurston's broader assertions.

\section*{Acknowledgements}

We would like to thank S.T. Joshi, D. Tsang, J. Read, I. Booth and C. Lackey for their insight on this project. 

for more information visit www.titaniumphysics.com

\bibliography{lovecraft}

\bibliographystyle{plain}

\end{document}